# Demonstration of Time-Reversal in Indoor Ultra-Wideband Communication: Time Domain Measurement

A.Khaleghi, G. El Zein, I.Naqvi
*Institut d'Electronique et de Télécommunications de Rennes (IETR)*
*Ali.khaleghi@insa-rennes.fr, Ghais.el-zein@insa-rennes.fr, Ijaz-haider.naqvi@ens.insa-rennes.fr*

***Abstract***- *Using time domain measurements, we assess the feasibility of time-reversal technique in ultra-wideband (UWB) communication. A typical indoor propagation channel is selected for the exploration. The channel response between receive and transmit antenna pairs is measured using time domain equipments which include an arbitrary wave generator (AWG) and a digital storage oscilloscope (DSO). The time-reversed version of the channel response is constructed with AWG and re-transmitted in the channel. The equivalent time reversed channel response is recorded. The properties of the time reversal technique in the line of sight (LOS) co-polar and cross-polar scenarios are measured.*

## 1. Introduction

Time-reversal (TR) has been proposed and studied for acoustical imaging, electromagnetic imaging, underwater acoustic communication, and very recently for wireless communication [1-6]. TR is a technique to focus spatially and compress temporally broadband signals in rich scattering environment. In this technique, the time reversed version of the channel response between transmitter and receiver is used as a pre-filter for data transmission. This has several advantages for UWB communication. The equivalent TR channel response is compressed in time and has a very short effective length, thus, the complex task of estimating a large number of taps at the receiver is greatly reduced. This implies low cost receivers. Furthermore, due to the considerable focusing gain, better signal to noise ratio or equally higher data rate can be achieved. It is also possible to increase the communication range by keeping FCC spectral mask limit for UWB communication. Another feature of TR is spatial focusing that can reduce greatly the co-channel interference in multi-cell systems.

The feasibility of TR in UWB has been demonstrated in [5]. However, the strategy involves indirect measurement of UWB channel in the frequency domain and then the data analysis in the time domain. In [6] the channel sounding is conducted using time-domain measurement but TR performance is demonstrated by the post-processing of the channel response. Both these approaches, give an ideal prediction of TR performance.

For the first time in this context, we have acquired not only UWB channel response in time-domain using a sub-nanosecond impulse and a high performance DSO, but also we have generated the time-reversed version of the channel response using an AWG. Time reversal is activated in two LOS channels in IETR laboratory environment. The characteristics of the equivalent TR channel (focusing gain, temporal compression and temporal sidelobes) are investigated.

## 2. Time-reversal

Considering the transmitter-receiver pair, TR uses time reversed response of the channel as the transmitter pre-filter. Denote the channel impulse response by $h(r_0,\tau)$, where $r_0$ is the receiver location and $\tau$ is the delay variable. By applying TR technique, the effective channel response to any location $r_0$ is thus given by

$$S(r_0,\tau) = X_{tr}(r_0,\tau) * h(r_0,\tau) + n_1(\tau) \qquad (1)$$

where $*$ denotes the convolution operation, $n_1(\tau)$ is the noise component at the TR channel and $X_{tr}$ is given by

$$X_{tr}(r_0,\tau) = A\,\hat{h}(r_0,-\tau) + n_2(\tau) \qquad (2)$$

This work was supported by ANR- MIRTEC. The authors are with IETR- UMR CNRS 6164- INSA

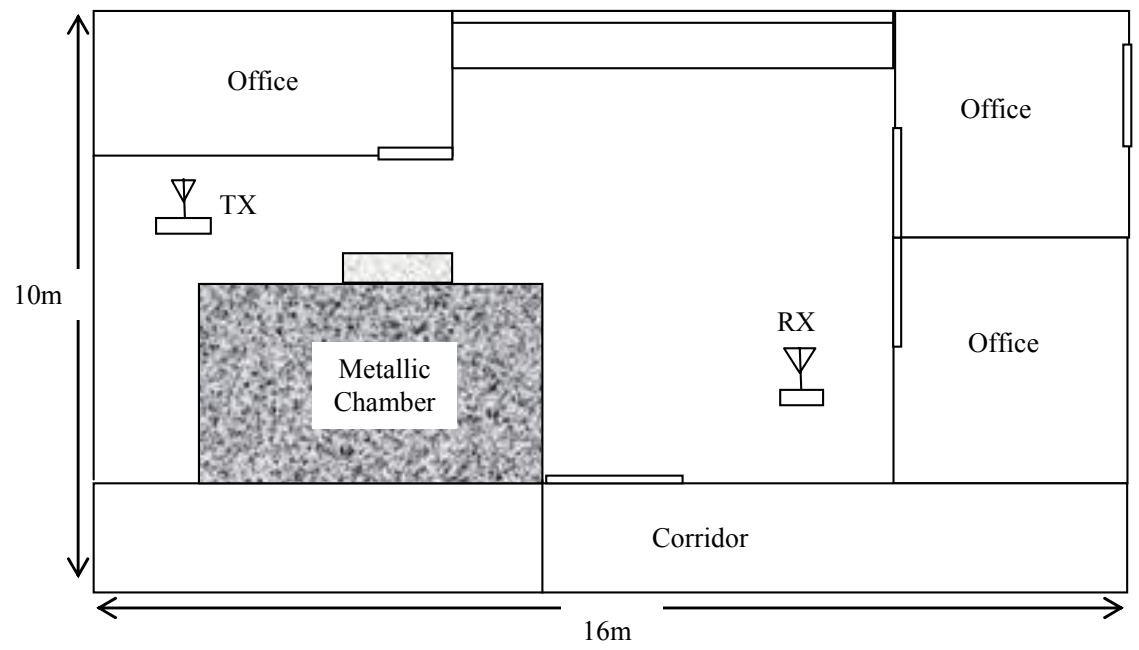

Fig.1 Measurement environment layout

where $\hat{h}(r_0,-\tau)$ is the time reversed version of the measured channel response that is truncated and then resampled in order constructed with AWG. $A$ is the amplification term to supply constant transmitted power ($P_o$) and $n_2(\tau)$ is the noise components of the measured channel. We assume the noise signal for each path component as an independent additive Gaussian variants with zero mean and standard deviation σ. Thus, the influence of the noise can be alleviated by a simple averaging process over multiple measures of the channel response. Therefore, the equivalent TR channel response (1) is expressed as

$$S(r_0,\tau) = A\hat{h}(r_0,-\tau) * h(r_0,\tau) + n_1(\tau) \approx A R_{hh}(\tau) + n_1(\tau) \quad (3)$$

where $R_{hh}(\tau)$ is the autocorrelation function of the channel response that peaks at $\tau$ =0. The amount of the signal peak depends on the energy of the truncated signal. We define the focusing gain (F.G) as the ratio of the strongest tap power in TR channel to the strongest tap power of the direct channel at location $r_0$

$$F.G = \max\left|S_{tr}(r_0,\tau)\right|^2 / \max\left|h(r_0,\tau)\right|^2 \quad (4)$$

where we have assumed that the transmitted average power in the direct and TR channels are both normalized to, $P_o$.

## 3. Experimental setup

To demonstrate the feasibility of time reversal UWB in indoor propagation environment a set of measurements are conducted in a modern laboratory building of IETR having the plan shown in Fig.1. All rooms are furnished by office equipments: tables, PCs and chairs. Moreover, there is a large metallic chamber in the environment, which probably increases the wave reflections in the environment.

We use two wideband conical monopole antennas (CMA) for signal transmission and reception. These antennas provide good impedance matching (return loss< -10dB) for the frequency range 0.7-8GHz. The radiation patterns of both antennas are almost uniform in the azimuth plane and are vertically polarized. The antennas are installed 1.5m above the ground and the separation between the antennas is 4.5m.

The measurement setup is illustrated in Fig.2. A Tektronix AWG 7052 is used for waveform generation. The maximum sampling rate of the system is 5GS/s and it is capable of generating signals with a -6dB bandwidth of 2.7GHz in the direct output mode. The output signal can take the amplitude of 1 $V_{p\text{-}p}$ with 50Ω output impedance. An impulse shaped waveform with minimum possible rise time is generated with AWG (see Fig.3). For our experiments the rise time is 230ps, this gives a wideband signal spectrum in the frequency range within DC to 2.7 GHz. The waveform directly feeds the transmitting antenna.

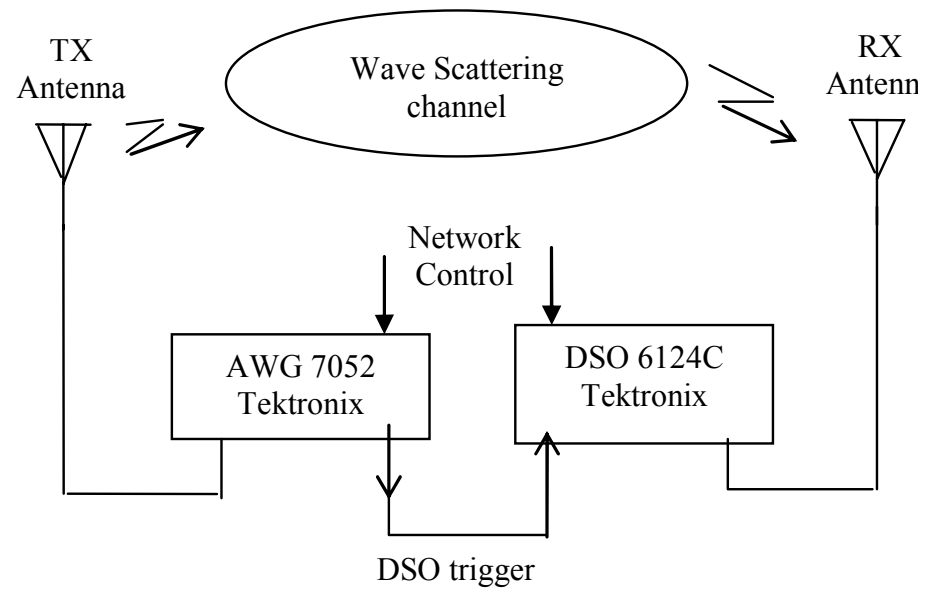

Fig.2 Experimental setup

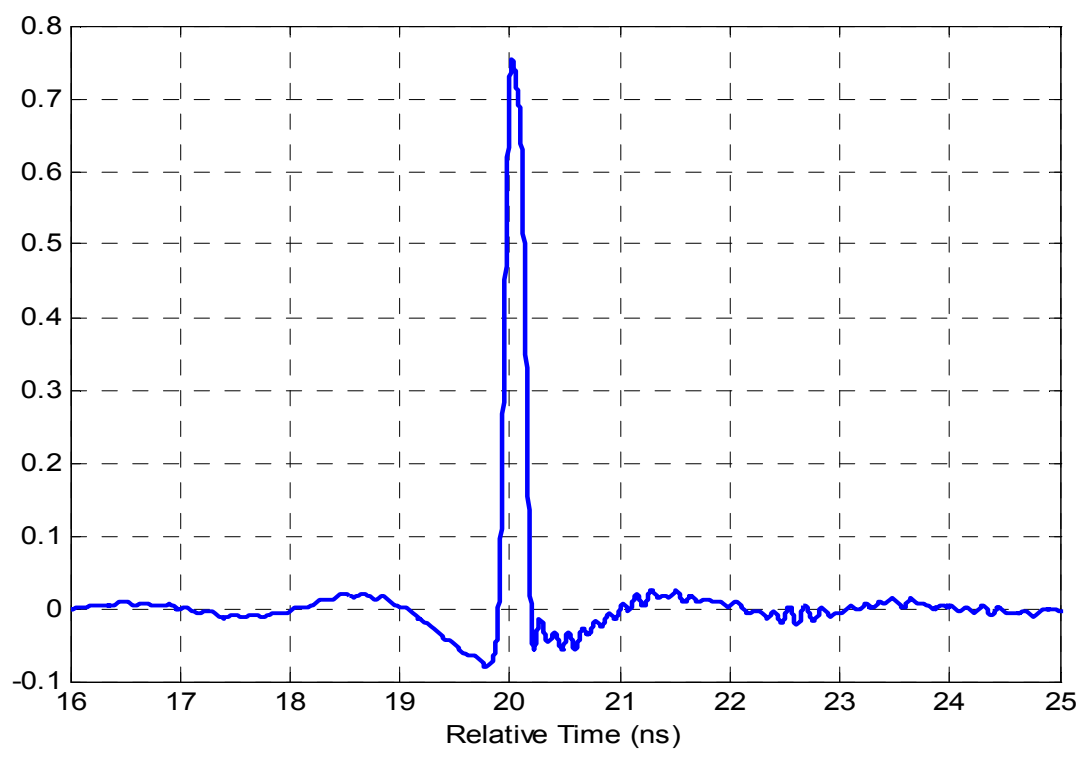

Fig.3 Impulse signal used for channel sounding

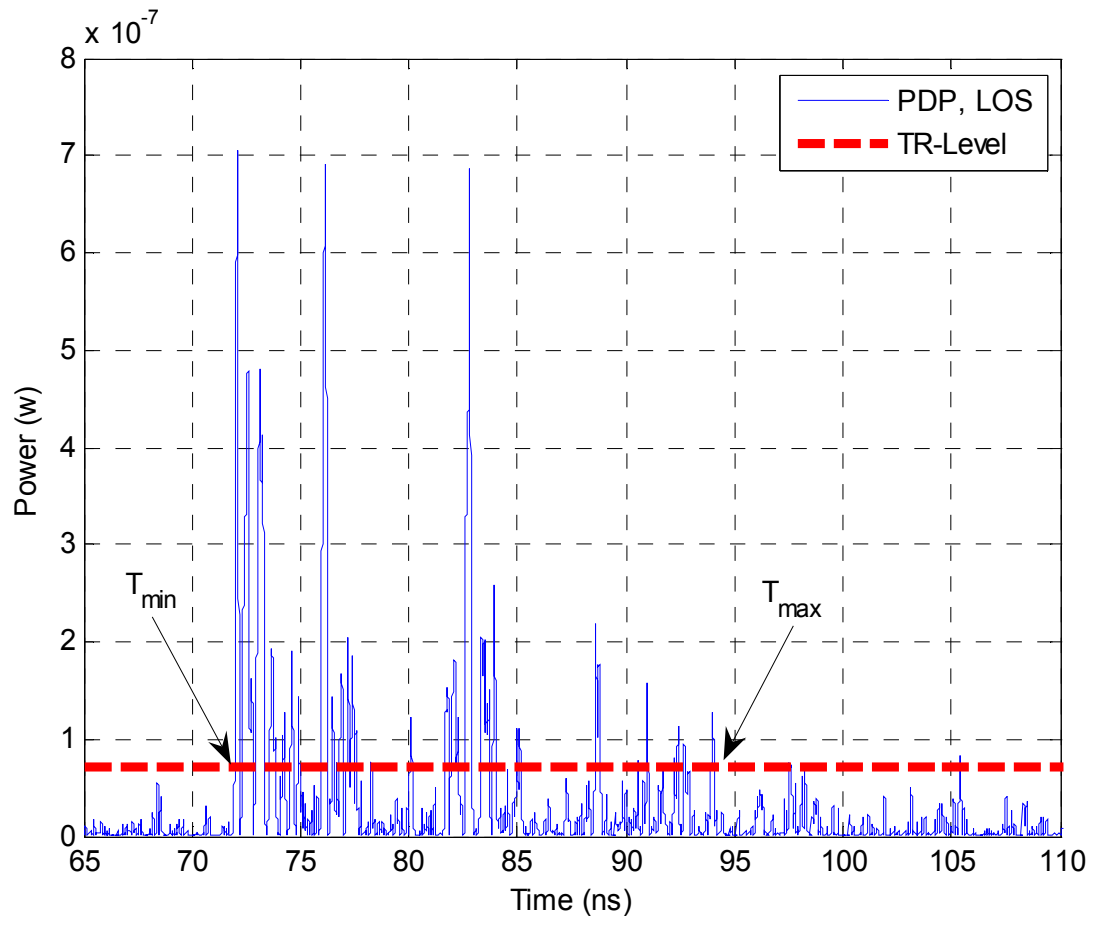

Fig.4 Measured instantaneous PDP in LOS channel

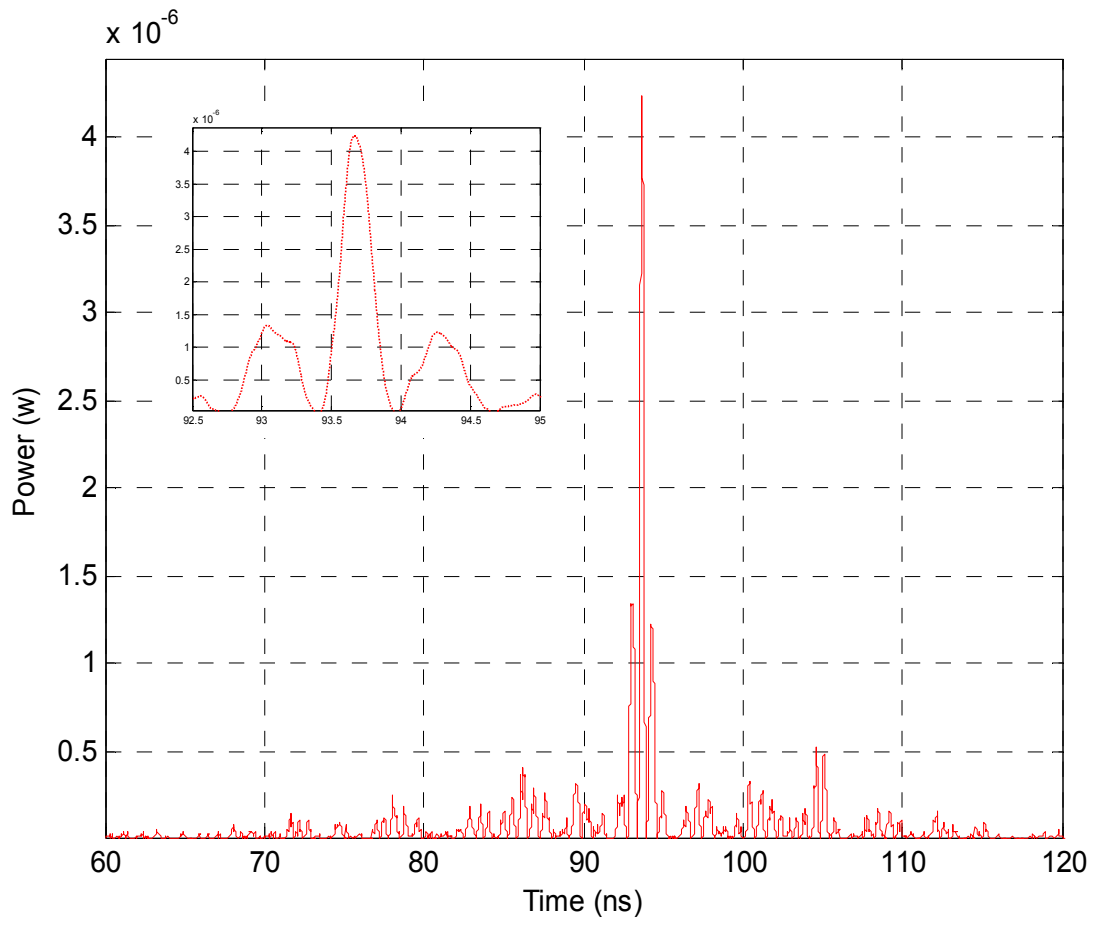

Fig.5 Measured time reversal response in LOS channel

The antennas act as filters and limit the signal bandwidth in their application range, thus, using the above mentioned antennas the transmitted signal spectrum is limited in the frequency range within 0.7-2.7GHz, and thus offers 2GHz bandwidth.

The received signal at RX antenna position is directly sampled and stored using a high performance Tektronix DSO 6124C, with 12GHz analog bandwidth and the sampling rate of 40GS/s. Average acquisition mode is used to reduce the effects of the additive noise in the measured signal samples. For instance, the channel is measured 128 times and then is averaged. The acquisition time for a total channel length of 400ns (8k signal samples) is almost 1sec. Therefore, the channel must be static during this short time.

To study the characteristics of the TR in indoor propagation channel two scenarios in LOS with the antennas in co-polar and cross-polar are realized. In co-polar scenario both of the antennas are vertically polarized but for cross-polar case the transmitting antenna is horizontally oriented. This will avoid the direct coupling between the antennas and the influence of the LOS propagation will be mitigated.

### 3.1. LOS (Co-polar) scenario

Fig.4 shows the measured instantaneous power delay profile (PDP) of the co-polar channel. As shown, due to the wave reflections and diffractions from the propagation environment, the received waveform is time delayed. Major parts of the multipath components (taps with more than 10 percent of the maximum tap power) are distributed within 22ns. The received signal level is reduced due to the free space loss, antenna impedance mismatching and the multipath propagation the antennas, can slightly modify the TR performance.

Fig.5 shows the measured instantaneous PDP of the TR channel. As shown, the received signal is compressed in time and has a very short effective length, thus, the signal detection process is simple. The measured F.G of the time reversal system is 7.8dB. In other words, for identical transmitted powers, the maximum peak power of TR channel is 7.8dB more than the maximum peak power in direct channel (see Fig.4).

It can be shown that the received energy of the signal in TR channel is 4dB more than the direct channel. Two reasons are distinguished for this observation: 1) the multipath components combined coherently 2) the power spectrum of the TR waveform is more adapted to the operating frequency band of the antennas.

Measurements show that the temporal sidelobes of the TR channel in co-polar mode are 5.3dB less than the main lobe level.

### 3.2. LOS (Cross-polar) channel

In cross-polar mode the transmitting antenna is horizontally polarized. This will avoid the direct coupling between the antennas, thus the strongest path components are generated by the wave reflections from the environment. The channel response for the transmitted impulse waveform (Fig.3) is measured. The PDP of this channel is shown in Fig.6. The delay spread of cross-polar channel is identical to the co-polar and the most important components are dispersed in a delay of 25ns. The received average power is 5.7dB lesser than the co-polar channel.

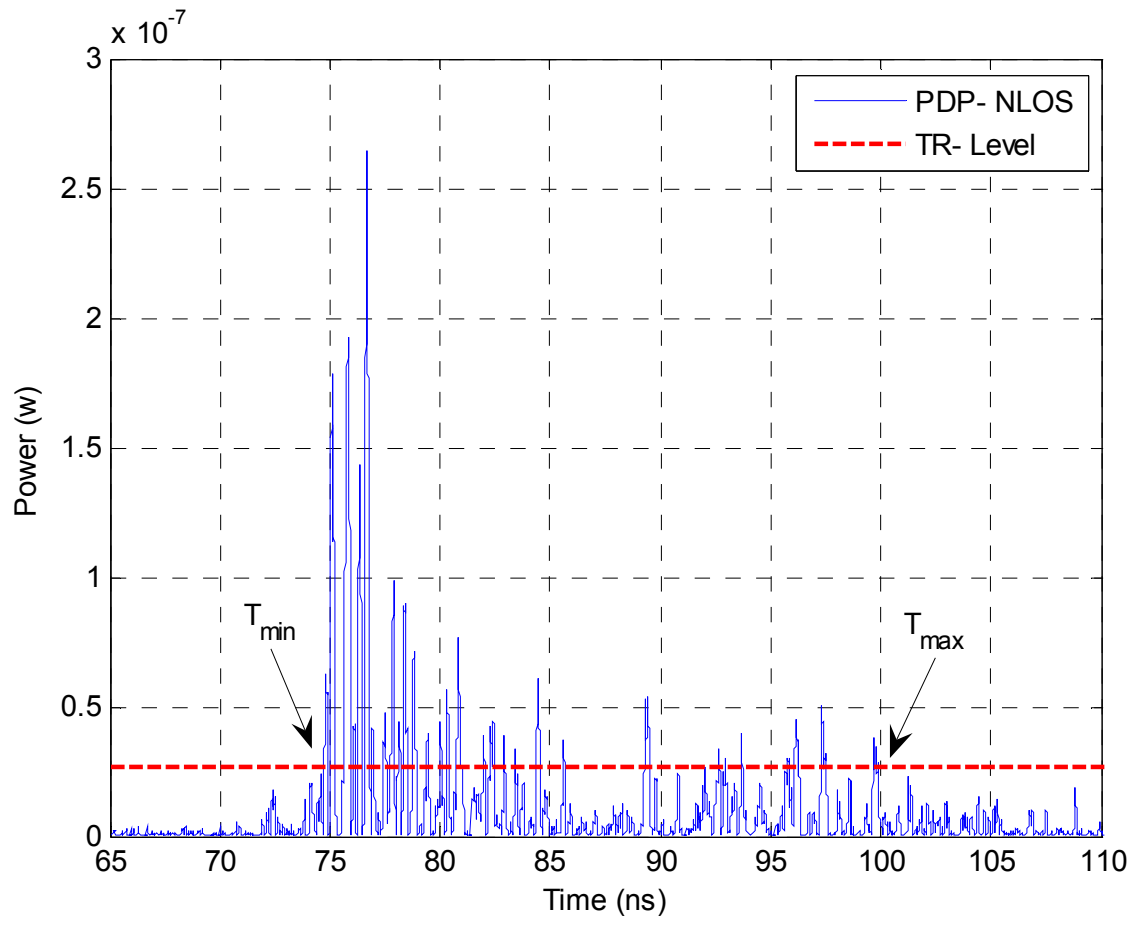

Fig.6 Measured instantaneous PDP in NLOS channel

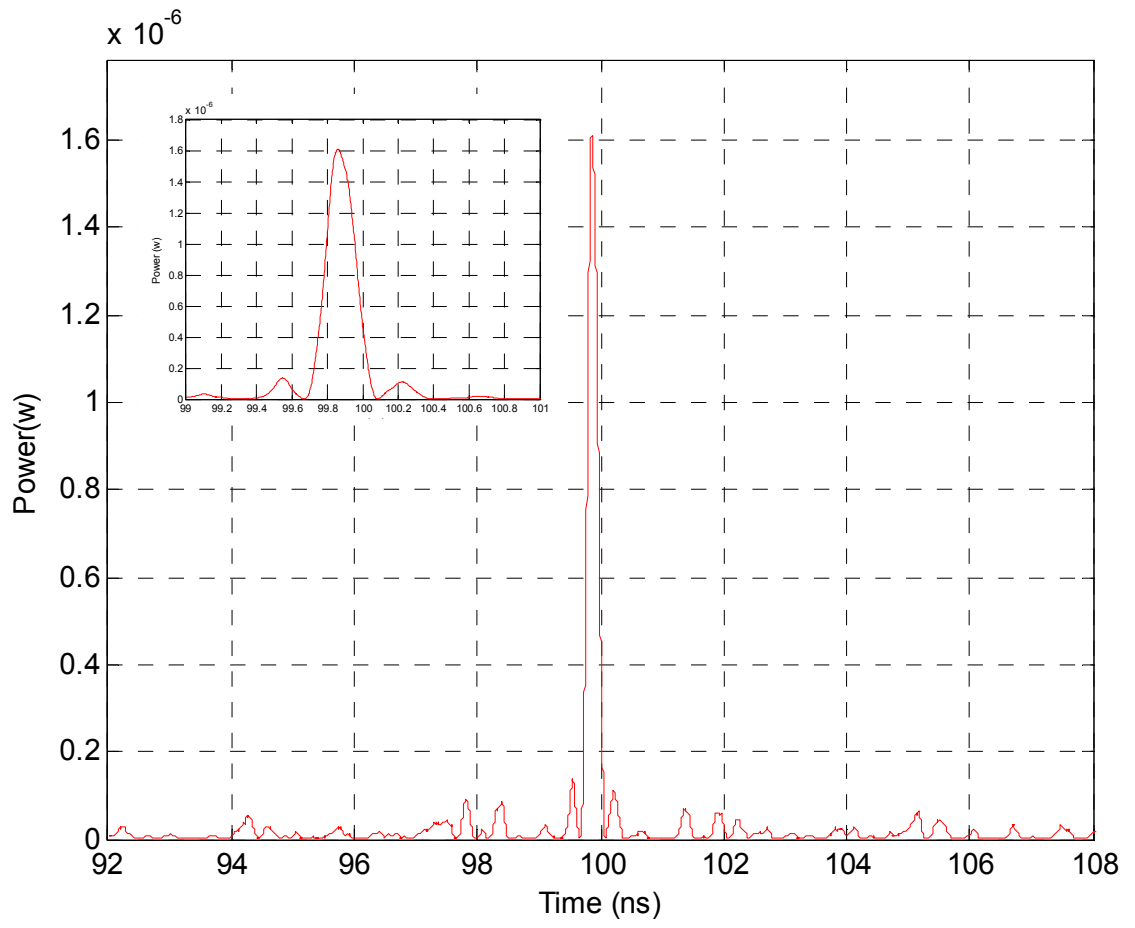

Fig.7 Measured time reversal response in NLOS channel

The time reversed version of the channel response in the specified time interval (see fig.6) is regenerated with AWG and retransmitted in the environment. Fig.7 shows the equivalent PDP of the TR channel. The measured F.G is 7.8dB. The recorded energy of the signal in TR channel is increased by 1.2dB. However, the temporal sidelobes are 11.5dB less than the main peak power.

Compared to the co-polar scenario, the temporal sidelobes are greatly reduced. Similar result is previously predicted for the LOS and non-LOS scenarios [5].

## 4. Conclusion

Time domain measurement is used to validate the feasibility of UWB time reversal in indoor environment. The performance of time reversal for LOS (co-polar and cross-polar) scenarios is measured. Temporal focusing property of the technique is shown and focusing gain of 7.8 dB is measured for both channels. Measurements indicate that the temporal sidelobes in co-polar channel are greater than cross-polar. An increment in the received signal energy of TR system is also identified.